\documentclass[letterpaper,10pt]{article} 

\usepackage{osameet3} 

\usepackage{amsmath,amssymb}
\usepackage[pdftex,colorlinks=true,bookmarks=false,citecolor=blue,urlcolor=blue]{hyperref} 
\usepackage{braket}

\newcommand{\bq}{\begin{equation}} \newcommand{\eq}{\end{equation}}
\newcommand{\bqali}{\bq\begin{aligned}}
\newcommand{\eqali}{\end{aligned}\eq}

\newcommand\rC{r_\text{\tiny C}}

\begin{document}

\title{Current tests of collapse models:
How far can we push the limits of quantum mechanics?
}

\author{Matteo Carlesso and Angelo Bassi}
\address{Department of Physics, University of Trieste, Strada Costiera 11, 34151 Trieste, Italy\\Istituto Nazionale di Fisica Nucleare, Trieste Section, Via Valerio 2, 34127 Trieste, Italy}
\email{matteo.carlesso@ts.infn.it}

\begin{abstract}
Collapse models implement a progressive loss of quantum coherence when the mass and the complexity of quantum systems increase. We will review such models and the current attempts to test their predicted loss of quantum coherence.
\end{abstract}



Collapse models describe the transition from the micro-world, well described by quantum mechanics, to the macro-world, where systems are never observed in superpositions. They modify the Schroedinger equation including the wave-function collapse in the dynamics  \cite{Bassi:2003aa}. Such a unified evolution is obtained by adding stochastic and non-linear terms. This solves the quantum measurements problem: the collapse is embedded in the dynamics and it does not only occur in the measurement processes.
For microscopic systems, collapse models induce negligible effects, confirming the well-known and experimentally verified results about quantum mechanics. Conersely, when the size and the complexity of the system increase moving towards the macroscopic realm, a built-in amplification mechanism ensures that the collapse is effective. It this way, the collapse suppresses the quantum superpositions and the systems behave classically \cite{Bassi:2003aa}. The quantum-to-classical transition is thus explained without run into paradoxes like the famous Schroedinger's cat.

The most studied among collapse models is the Continuous Spontaneous Localization (CSL) model, which describe the collapse as occurring continuously in time, and it can be considered as a valid alternative to quantum mechanics. 
Two free phenomenological parameters characterize the model. These are the collapse rate $\lambda$ and the correlation distance of the collapse noise $\rC$. During the years, different proposals for these values were presented. Ghirardi, Rimini and Weber (GRW) originally set \cite{Bassi:2003aa} $\lambda=10^{-16}$\,s$^{-1}$ and $\rC=10^{-7}$\,m. Later, Adler suggested different values  \cite{Adler:2007ab} namely $\rC=10^{-7}$\,m with $\lambda=10^{-8\pm2}$\,s$^{-1}$ and $\rC=10^{-6}$\,m with $\lambda=10^{-6\pm2}$\,s$^{-1}$. A consensus so far has been not reached, but, as the model is phenomenological, the values of the parameters must be eventually determined by experimental evidences. 

The collapse effects are made proportional to the mass of the system. Thus, the amplification mechanism is automatically implemented, and one can focus on the contributions coming from the nucleons and neglect those from the electrons. The collapse dynamics gives different predictions from those given by quantum mechanics. The comparison of these results with the experimental results will constrain the CSL parameter space, as we will now below.
Although the development of dedicated experiments has started only very recently \cite{Vinante:2017aa}, the comparison of the predictions of the CSL model with the available experimental data has already set strong bounds on the collapse parameters. The experiments that have been exploited so far can be distinguished into families: the interferometric and non-interferometric tests.

 \begin{figure}[h!]
\begin{center}
\includegraphics[width=0.5\linewidth]{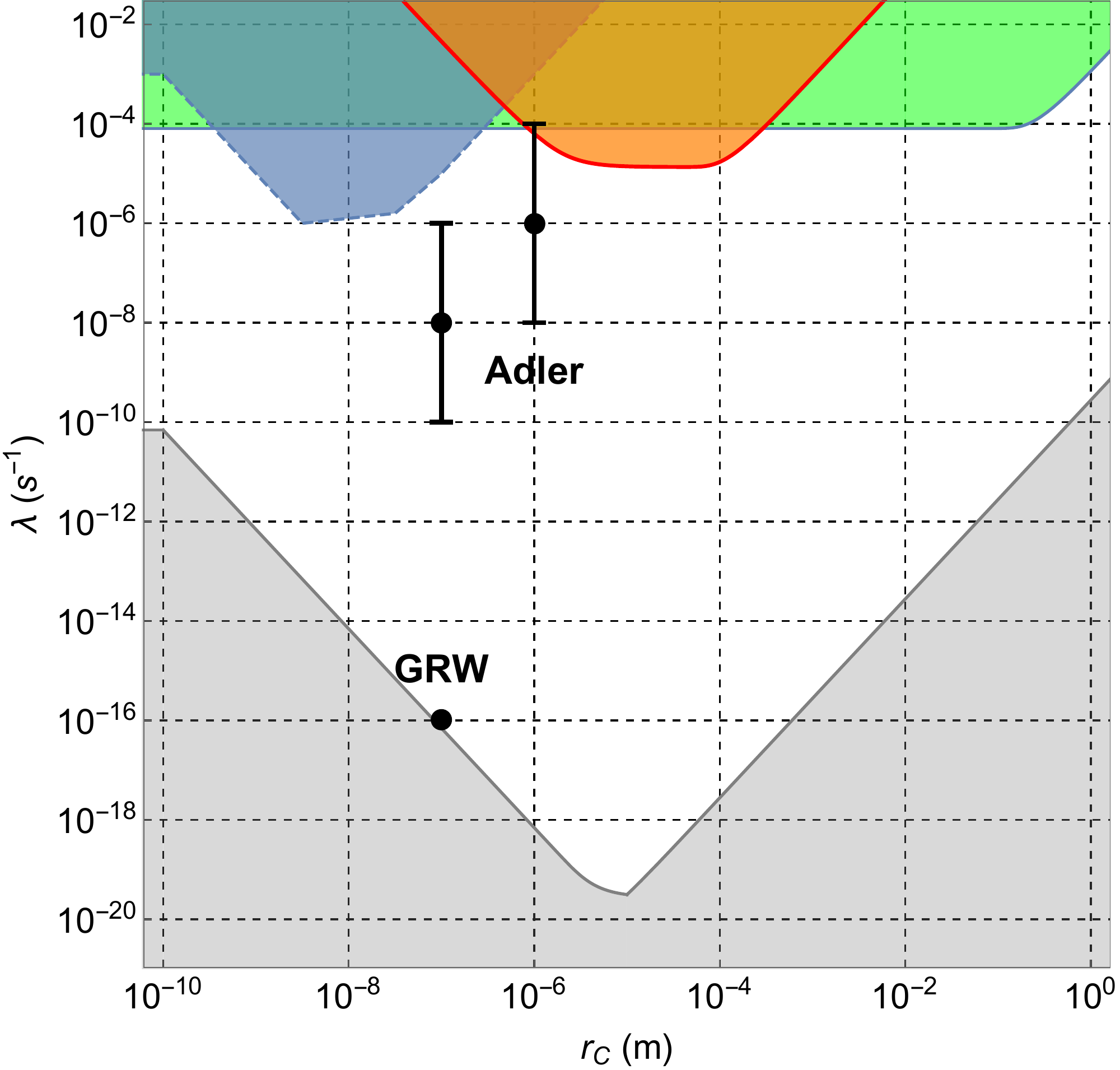}\includegraphics[width=0.5\linewidth]{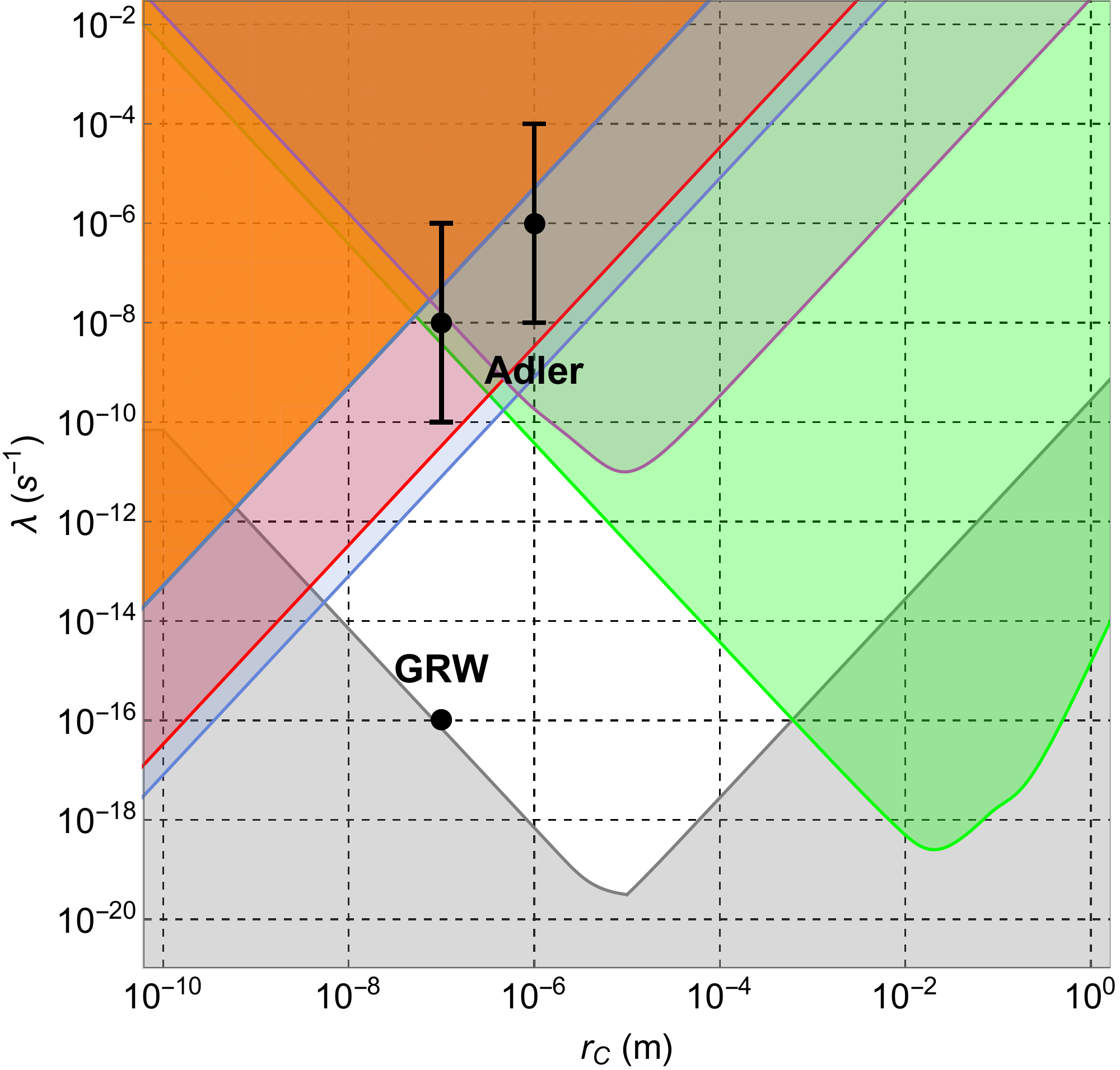}
\end{center}
\caption{Exclusion plots for the CSL parameters with respect to the GRW's and Adler's theoretically proposed values\cite{Bassi:2003aa,Adler:2007ab}. \textit{Left panel} - Excluded regions from interferometric experiments: molecular interferometry\cite{Eibenberger:2013aa,Toros:2017aa} (blue area), atom interferometry \cite{Kovachy:2015aa} (green area) and experiment with entangled diamonds \cite{Lee:2011aa} (orange area). \textit{Right panel} - Excluded regions from non-interferometric experiments: LISA Pathfinder \cite{Carlesso:2016ac,Armano:2018aa} (green area), cold atoms \cite{Kovachy:2015ab} (orange area), phonon excitations in crystals~\cite{Adler:2018aa} (red area), X-ray measurements~\cite{Aalseth:1999aa} (blue area) and nanomechanical cantilever \cite{Vinante:2017aa}. We report with the grey area the region excluded based on theoretical arguments \cite{Toros:2017aa}. }
\label{fig1}
\end{figure}


In interferometric experiments, we have those where a quantum superposition is created and eventually probed. Since collapse models suppress quantum coherences and localize the state of the system in space, these experiments are the most natural way to test collapse theories. The general procedure is to detect the interference pattern and quantify its reduction. This will place an upper bound on the maximum value of the parameter $\lambda$ given a value for $\rC$.
Examples of these experiments are atom \cite{Kovachy:2015aa} and molecular\cite{Eibenberger:2013aa,Toros:2017aa} interferometry and entanglement experiment with diamonds \cite{Lee:2011aa}. The corresponding upper bounds are shown in Figure \ref{fig1}.

Following the same point of view, one can also set a lower bound starting from theoretical considerations\cite{Toros:2017aa}. Indeed, the collapse must guarantee that macroscopic objects actually collapse. This sets a minimum value for $\lambda$.
The left panel of Figure \ref{fig1} shows the upper bounds obtained from interferometric experiments, compared with the GRW's and Adler's theoretically proposed values of the parameters and the lower theoretical bound.


The strongest bounds to the CSL model are, so far, not those from interferometric experiments, but from non-interferometric ones. These experiments, instead of involving the creation of a quantum superposition, focus on a side effect of collapse models. Indeed, CSL predicts a Brownian motion of the center of mass of the system, which can be detected by the extremely fine-tuned non-interferometric techniques \cite{Bahrami:2014aa, Armano:2018aa}. 
Since no superposition is created, such experiments can involve also truly macroscopical systems, where collapse effects are more effective due to the amplification mechanism.
These experiments involve cold atoms~\cite{Kovachy:2015ab}, optomechanical systems~\cite{Vinante:2006aa,Carlesso:2016ac,Vinante:2017aa,Armano:2018aa}, X-ray measurements~\cite{Aalseth:1999aa} and phonon excitations in crystals~\cite{Adler:2018aa}.
The right panel of Figure \ref{fig1} shows the upper bounds that can be inferred from the existing non-interferometric tests.

The authors acknowledge support from the University of Trieste (FRA 2016), INFN, the COST Action QTSpace (CA15220) and the H2020 FET project TEQ (grant n. 766900).

\end{document}